\documentclass{eptcs}
\providecommand{\event}{CompMod 2009} 
\providecommand{\volume}{??}
\providecommand{\anno}{20??}
\providecommand{\firstpage}{1}
\providecommand{\eid}{??}

\usepackage{breakurl}        
\usepackage{graphics}
\usepackage{graphicx}
\usepackage{epsfig}
\usepackage{amssymb}
\usepackage{amsthm}

\newcommand{\virg}{``}

\title{A study on the combined interplay between\\ stochastic fluctuations and the number of flagella\\ in bacterial chemotaxis}

\author{Daniela Besozzi$^a$ \quad Paolo Cazzaniga$^b$  \quad Matteo Dugo$^a$ \\Dario Pescini$^b$  \quad Giancarlo Mauri$^b$\\
\institute{$^a$Universit\`{a} degli Studi di Milano\\ Dipartimento di Informatica e Comunicazione\\ Via Comelico 39, 20135 Milano, Italy}
\email{besozzi@dico.unimi.it, matteo.dugo@studenti.unimi.it}\\
\institute{$^b$Universit\`{a} degli Studi di Milano-Bicocca\\ Dipartimento di Informatica, Sistemistica e Comunicazione\\ Viale Sarca 336, 20126 Milano, Italy}
\email{cazzaniga/pescini/mauri@disco.unimib.it}
}

\def\titlerunning{Interplay of stochastic fluctuations and number of flagella in bacterial chemotaxis}
\def\authorrunning{D. Besozzi et al.}
\begin{document}
\maketitle

\begin{abstract}
The chemotactic pathway allows bacteria to respond and adapt to environmental changes, by tuning the tumbling and running motions that are due to clockwise and counterclockwise
rotations of their flagella. The pathway is tightly regulated by feedback mechanisms governed by the phosphorylation
and methylation of several proteins. In this paper, we present a detailed mechanistic model for chemotaxis, that considers all of its transmembrane and cytoplasmic components, and their mutual interactions. Stochastic simulations of the dynamics of a pivotal protein, CheYp, are performed by means of tau leaping algorithm. This approach is then used to investigate the interplay between the stochastic fluctuations of CheYp amount and the number of cellular flagella. Our results suggest that the combination of these factors might represent a relevant component for chemotaxis. Moreover, we study the pathway under various conditions, such as different methylation levels and ligand amounts, in order to test its adaptation response. Some issues for future work are finally discussed.
\end{abstract}

\section{Introduction}

Recent experimental investigations at the single-cell level \cite{Elowitz_stoch} have evidenced the presence of biological noise, due to the inherently stochastic interactions between those molecular species that occur in low amounts inside the cell. Standard modeling approaches based on ordinary differential equations cannot effectively capture the effects of biological random processes, which are able to lead the cell to different states (e.g., bistability). In the last years, indeed, many algorithms that perform stochastic simulations of biochemical reaction systems have proved their intrinsic suitability for reproducing the dynamics of many cellular processes (see, e.g., \cite{Tur04} and references therein). For instance, the stochastic simulation algorithm (SSA) \cite{SSA} is able to provide an exact realization of the system dynamics and to account for random fluctuations in the temporal evolution of molecules; however, it can be very slow for those systems where the number of reactions -- or even the amount of a few molecular species -- is large, as it is frequently the case for complex cellular processes involving many species and many interactions among them. In order to overcome this drawback, a faster and reliable stochastic simulation algorithm, called tau leaping \cite{Gill06}, has been recently proposed.
Tau leaping represents an efficient tool for the modeling of stochastic processes in individual cells (see, e.g., \cite{cAMP}), as it can easily handle very detailed and mechanistic descriptions of all possible molecular interactions or molecule modifications, which can bring about a huge increase in the number of reactions and molecular species (e.g., phosphorylation or methylation states of proteins). In this work we exploit the efficiency of tau leaping to simulate the dynamics of bacterial chemotaxis, in order to generate stochastic time series of a pivotal chemotactic protein, which will be then analyzed with respect to the number of flagella in bacterial cells.

Chemotaxis is an efficient signal transduction pathway which allows bacterial
cells to move directionally, in response to specific attractants or repellents occurring in their surroundings. The pathway consists of several transmembrane and cytoplasmic proteins acting as signal sensors and response regulators \cite{Jurica}, which rule the reversal of the flagellar motor (governed by the phosphorylation and dephosphorylation of a key protein, CheY). This process induces a switch between running and tumbling movements, with a frequency that allows a temporal sampling (through random walks) of homogeneous environments. Anyway, in the presence of a gradient of attractants or repellents, the bacteria are able to respond quickly by reducing the frequency of flagellar reversal between clockwise and counterclockwise rotations, which cause a longer running motion in a biased direction. The frequency of switching is then reset to the random walk level if the concentration of the external ligands remains constant in time. At the molecular scale, this adaptation property is implemented by the coordinated action of methyltransferase and methylesterase proteins acting on the transmembrane receptors.

The genetic regulation and biochemical functions of the proteins involved in chemotaxis are well known, and several models have already been proposed to study their complex interplay as well as the robustness of this system \cite{robust,Spiro,Morton,Levin,Lipkow,RaoPLOS}. In the model we present hereby, we consider very detailed protein-protein interactions for the chemotactic pathway in \emph{E. coli}, in response to attractant molecules, which sum up to 62 biochemical reactions and  32 molecular species. The temporal evolution of the phosphorylated form of CheY (CheYp) is investigated under different conditions, such as the deletion of other proteins involved in the pathway, the addition of distinct amounts of external ligand, and the effect of different methylation states.

The results obtained through stochastic simulations of this model are then used to propose an analysis on the interplay between the stochastic fluctuations of CheYp and the number of cellular flagella, which occur in a few units in the individual bacterium (around half a dozen in \emph{E. coli}). The aim of this analysis is to devise the mean time periods during which the cell either performs a running or a tumbling motion, considering both the coordination of flagella and the randomness that is intrinsic in the chemotactic pathway. Indeed, experimental observations show that the running motion requires all flagella to be simultaneously synchronized in the counterclockwise rotation, which occurs when CheYp is not interacting with the proteins regulating the flagellar motor; on the contrary, when at least one flagellum is not coordinated with the others, then the bacterium performs a tumbling movement. To distinguish between these two states, we will assume that the cell is sensitive to a threshold level of CheYp, that is evaluated as the mean value of CheYp at steady state. Because of stochastic fluctuations, the amount of CheYp will randomly switch from below to above this value, thus reversing the rotation from counterclockwise to clockwise rotations of each flagellum. Therefore, the original contribution of our work consists in linking the synchronization of all flagella to the stochastic fluctuations of CheYp, as the core component that stands at the basis of chemotactic motions. To this aim, we define a procedure to identify the synchronization of rotations of all flagella, and we use it to compare the mean time intervals of running and tumbling motions -- as well as of the adaptation times after ligand addition -- according to a varying number of flagella.

The paper is structured as follows. In Section \ref{sec:model} we give a description of the bacterial chemotactic pathway, and present the mechanistic model of this system. In Section \ref{sec:sim_dyn}, after a brief explanation of the functioning of tau leaping, we show the results of simulations for the temporal evolution of the pivotal protein in chemotaxis (CheYp), that have been obtained by using the model previously defined. Then, in Section \ref{sec:interplay} we exploit the outcome of stochastic simulations to study the relations between the fluctuations of CheYp and the number of flagella occurring in a bacterial cell. We conclude the paper with some topics for future extensions of our work.

\section{The modeling of bacterial chemotaxis}\label{sec:model}

In this section we present the chemotaxis signaling pathway and define the mechanistic model that describes the molecular interactions therein involved.

\subsection{Bacterial chemotaxis}\label{ssec:bio}

Chemotaxis is a signal transduction pathway that allows swimming bacteria to perform biased movements in ever-changing environments, by efficiently sensing concentration gradients of beneficial or toxic chemicals in their immediate proximity. The binding of ligand molecules triggers an event cascade involving several transmembrane and cytoplasmic proteins, which eventually affects the concentration of a pivotal response regulator, CheY. This protein rapidly diffuses inside the cell and interacts with the proteins of the flagellar motors, thus inducing clockwise (CW) and counterclockwise (CCW) rotation of each flagellum. When flagella are turning CW, they are uncoordinated and the bacterium performs a \emph{tumbling} movement, while if they are all turning CCW, they form a bundle and get coordinated, thus allowing the cell to swim directionally (the so-called \emph{running} movement). In a homogeneous environment, bacteria perform a
temporal sampling of their surroundings by moving with a random walk, that is caused by a high switch frequency of the flagellar motors rotations, that alternate rapid tumblings with short runnings. In the presence of a ligand concentration gradient, instead, bacteria carry out directional swimming toward/against the attractants/repellents, by reducing the switch frequency of flagella rotations, that results in longer running movements. If the ligand concentration remains constant in time, then the switch frequency is reset to the prestimulus level, therefore realizing an \emph{adaptation} of the chemotactic response to the change in ligand concentration. In what follows, we consider the chemosensory system of \emph{E. coli} bacteria, in response to attractant chemicals.

\begin{figure}
\centering
\includegraphics[height=5cm]{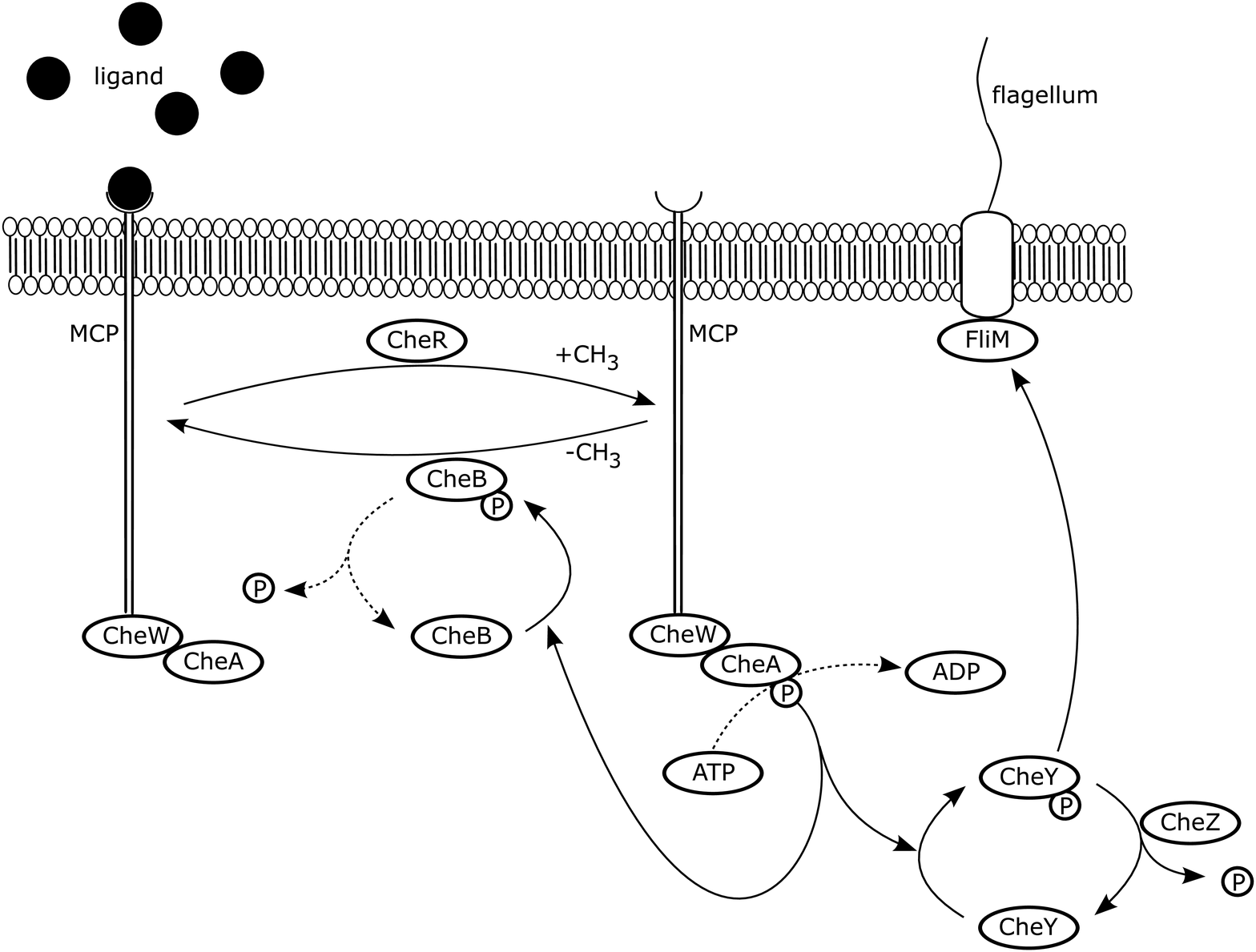}
\caption{Signal transduction pathway in bacterial chemotaxis: solid arrows indicate enzyme-catalyzed reactions, dashed arrows indicate autocatalysis; CH$_3$ denotes the methyl group, P the phosphoryl group (the dimensions of components are not scaled).} \label{fig:pathway}
\end{figure}

The chemotactic pathway, depicted in Figure \ref{fig:pathway}, has been well characterized from a molecular point of view \cite{Bren,Jurica,Wadhams_chem}. External signals are detected by transmembrane methyl-accepting proteins (MCPs), which are linked to cytoplasmic histidine protein kinases (CheA) by means of scaffold proteins (CheW). These three proteins constitute the sensor module (i.e. the receptor complexes) of the whole pathway; each protein occurs as a dimer in every receptor complex. The role of CheA is to transduce the presence of an external ligand toward the inside of the cell, by phosphorylating two cytoplasmic proteins, called CheY and CheB. The transfer of the phosphoryl group to these proteins is more probable -- that is, the activity of CheA is stronger -- in absence of external ligands. CheY and CheB compete for the binding to CheA, but the phosphotransfer to CheY is faster than to CheB \cite{Wadhams_chem}; this fact assures that the proper chemotactic response can be generated before the process of adaptation occurs, as explained hereafter. CheY is the response regulator protein which, after being phosphorylated, interacts with the proteins FliM of the flagellar motors, inducing the CW rotation of the flagellum and the tumbling movements (FliM is a key component of the processes that stand \emph{downstream} of the chemotaxis signaling, and therefore will not be explicitly included in our model; anyway, some considerations about its role within the model are discussed in Section \ref{concl}). In presence of external ligands, the activity of CheA is reduced: the concentrations of phosphorylated CheY diminishes, its interaction with the flagellar motors is reduced, the CCW rotation is switched on, and bacteria can perform longer running movements. The termination of this signal transduction process is mediated by another cytoplasmic protein, CheZ, which acts as an allosteric activator of CheY dephosphorylation. Concurrently to the processes involving CheY, the chemosensory system possesses an adaptation response which depends on the methylation level of the receptors. Methylation reactions are modulated by the coordinated interplay between proteins CheR and CheB. Up to 4-6 methyl groups are constantly transferred to the cytoplasmic domain of MCPs by the constitutively active methyltransferases CheR. On the other side, the demethylation of MCPs occurs by means of the phosphorylated form of the methylesterase CheB. The methylation state of MCPs also intervene on the regulation of CheA: when MCPs are highly methylated, CheA is more active; when MCPs are unmethylated, the activity of CheA is reduced. In the latter case, also the concentrations of phosphorylated CheB diminishes, and this in turn lets the methylation state of MCPs increase, with a consequent renewed activity of CheA, and so on through a continuous feedback control. Therefore, the cell is able to adapt to environmental changes and return to the random walk sampling when the concentration gradient of the attractant remains constant in time. This feedback mechanism also allows bacteria to widen the range of ligand concentration to which they can respond, making them very sensible to low environmental variations.

\subsection{A mechanistic model}\label{ssec:stochmod}

For the modeling of the chemotaxis pathway, we have considered detailed protein-protein interactions which sum up to a total of 62 reactions and 32 molecular species \cite{Dugo}. The initial amounts -- given as number of molecules, as reported in \cite{Shimizu} -- of the 7 elementary chemotactic proteins are the following:  4000 dimers of MCPs; 4000 dimers of CheW; 4000 dimers of CheA; 17000 monomers of CheY; 12000 monomers of CheZ; 200 monomers of CheR;
1700 monomers of CheB (plus a constant amount of 1.2 $\cdot 10^6$  ATP molecules that are needed for phosphorylation reactions). The initial amounts of all other molecular species appearing in the model
are equal to zero, as they are produced by mimicking the formation and dissociation of protein complexes, and by describing the phosphorylation/dephosphorylation of cytoplasmic proteins and the methylation/demethylation of MCPs, in both the conditions of presence and absence of external ligands.

Each reaction in the model is given in the form \virg reagents $\rightarrow$ products'', where the notation X + Y is used to represent a molecular interaction between species X and Y, while X::Y denotes that X and Y are chemically bound in the formation of a complex (see Table \ref{tab:model}). Note that only monomolecular or bimolecular reactions are here considered; the formation of complexes consisting of more than two species is performed stepwise. The phosphorylated form of species X, with X $\in \{$CheA, CheB, CheY$\}$, is denoted by Xp, while the methylation state of receptor MCP is denoted by MCP$^m$, for $m=0, \dots, 4$ (that is, five methylation states are considered).

The reactions describe the following molecular interactions:
\begin{itemize}
\item association of the three dimers (2MCP, 2CheW and 2CheA) constituting each ternary receptor complex (reactions 1-4);
\item binding and unbinding of ligand molecules to the
receptor complex in the five methylation states (reactions 28-32 and 33-37, respectively);
\item methylation and demethylation of MCPs, in absence and in presence of ligand molecules
(reactions 5-8 and 9-12, 38-41 and 42-45, respectively);
\item autophosphorylation of CheA in
the five methylation states of MCPs, in absence and in presence of
ligand molecules (reactions 13-17 and 46-50, respectively);
\item phosphotransfer to CheY in the five methylation states of
MCPs, in absence and in presence of ligand molecules (reactions 18-22
and 51-55, respectively);
\item phosphotransfer to CheB in the five methylation states of
MCPs, in absence and in presence of ligand molecules (reactions 23-27
and 56-60, respectively);
\item dephosphorylation of CheYp and CheBp (reactions 61-62).
\end{itemize}

According to literature, the ternary receptor complex 2MCP$^m$::2CheW::2CheA is assumed to be stable for the duration of the signal transduction process \cite{Sourjik}; moreover, the synthesis and degradation rates of all chemotactic proteins are assumed to occur at a much slower scale than the chemotactic response (hence, the reactions corresponding to these processes have not been included in the model).

\begin{table*}[!th]
\caption{The 62 reactions of the model of bacterial chemotaxis. The values of the corresponding stochastic constants are:
$c_{1} = 0.1, c_{2} = 0.01, c_{3} = 0.1, c_{4} = 0.02, c_{5} = 0.325, c_{6} = 0.29, c_{7} = 0.165, c_{8} = 0.05, c_{9} = 0.0044, c_{10} = 0.0175, c_{11} = 0.0306, c_{12} = 0.035, c_{13} = 5.0 \cdot 10^{-7}, c_{14} = 7.0 \cdot 10^{-6}, c_{15} = 2.8 \cdot 10^{-5}, c_{16} = 5.0 \cdot 10^{-5}, c_{17} = 6.8 \cdot 10^{-5}, c_{18} = 5.0 \cdot 10^{-4}, c_{19} = 0.0035, c_{20} = 0.014, c_{21} = 0.025, c_{22} = 0.0336, c_{23} = 2.0 \cdot 10^{-4}, c_{24} = 0.0014, c_{25} = 0.0056, c_{26} = 0.01, c_{27} = 0.0135, c_{28} = 0.6, c_{29} = 0.8, c_{30} = 1.0, c_{31} = 1.2, c_{32} = 1.4, c_{33} = 15.0, c_{34} = 15.0, c_{35} = 15.0, c_{36} = 15.0, c_{37} = 15.0, c_{38} = 4.0 \cdot 10^{-4}, c_{39} = 3.75 \cdot 10^{-4}, c_{40} = 3.5 \cdot 10^{-4}, c_{41} = 2.125 \cdot 10^{-4}, c_{42} = 6.0 \cdot 10^{-4}, c_{43} = 0.0044, c_{44} = 0.0175, c_{45} = 0.0343, c_{46} = 1.0 \cdot 10^{-8}, c_{47} = 5.0 \cdot 10^{-7}, c_{48} = 7.0 \cdot 10^{-6}, c_{49} = 2.8 \cdot 10^{-5}, c_{50} = 5.0 \cdot 10^{-5}, c_{51} = 1.0 \cdot 10^{-5}, c_{52} = 5.0 \cdot 10^{-4}, c_{53} = 0.0035, c_{54} = 0.014, c_{55} = 0.03, c_{56} = 1.0 \cdot 10^{-5}, c_{57} = 2.0 \cdot 10^{-4}, c_{58} = 0.0014, c_{59} = 0.0056, c_{60} = 0.0112, c_{61} = 0.0080, c_{62} = 1.0$}
\label{tab:model}
\begin{center}\begin{small}
\begin{tabular}{|l|l|l|l|}
  \hline
   & \textbf{ Reagents} & \textbf{ Products} & \textbf{Methyl. state}\\
  \hline
1  & 2MCP$^m$ + 2CheW & 2MCP$^m$::2CheW & $m=0$\\
2  & 2MCP$^m$::2CheW & 2MCP$^m$ + 2CheW & $m=0$\\
3  & 2MCP$^m$::2CheW + 2CheA & 2MCP$^m$::2CheW::2CheA &  $m=0$\\
4 &  2MCP$^m$::2CheW::2CheA & 2MCP$^m$::2CheW + 2CheA & $m=0$\\
5-8 & 2MCP$^m$::2CheW::2CheA + CheR  &  2MCP$^{m+1}$::2CheW::2CheA + CheR &  $m=0, \dots, 3$\\
9-12 & 2MCP$^m$::2CheW::2CheA + CheBp  &  2MCP$^{m-1}$::2CheW::2CheA + CheBp &  $m=1, \dots, 4$\\
13-17 &  2MCP$^m$::2CheW::2CheA + ATP & 2MCP$^m$::2CheW::2CheAp & $m=0, \dots, 4$\\
18-22 &  2MCP$^m$::2CheW::2CheAp + CheY & 2MCP$^m$::2CheW::2CheA + CheYp &  $m=0, \dots, 4$\\
23-27 &  2MCP$^m$::2CheW::2CheAp + CheB & 2MCP$^m$::2CheW::2CheA + CheBp & $m=0, \dots, 4$\\
28-32 & lig + 2MCP$^m$::2CheW::2CheA  &  lig::2MCP$^m$::2CheW::2CheA & $m=0, \dots, 4$ \\
33-37 & lig::2MCP$^m$::2CheW::2CheA & lig + 2MCP$^m$::2CheW::2CheA & $m=0, \dots, 4$\\
38-41 & lig::2MCP$^m$::2CheW::2CheA + CheR & lig::2MCP$^{m+1}$::2CheW::2CheA + CheR &  $m=0, \dots, 3$\\
42-45 & lig::2MCP$^m$::2CheW::2CheA + CheBp  &  lig::2MCP$^{m-1}$::2CheW::2CheA + CheBp & $m=1, \dots, 4$\\
46-50 & lig::2MCP$^m$::2CheW::2CheA + ATP  &  lig::2MCP$^m$::2CheW::2CheAp &  $m=0, \cdots, 4$\\
51-55 & lig::2MCP$^m$::2CheW::2CheAp + CheY  &  lig::2MCP$^m$::2CheW::2CheA + CheYp & $m=0, \dots, 4$\\
56-60 & lig::2MCP$^m$::2CheW::2CheAp + CheB  &  lig::2MCP$^m$::2CheW::2CheA + CheBp & $m=0, \dots, 4$\\
61 &  CheYp + CheZ & CheY + CheZ  &  \\
62 &  CheBp & CheB  &   \\
  \hline
\end{tabular}
\end{small}\end{center}
\end{table*}

A stochastic constant is associated to each reaction: it is needed to evaluate the probability of that reaction to occur when performing stochastic simulations, as explained in Section \ref{ssec:tau}. The stochastic constants used for the simulations shown in Section \ref{ssec:sim_results} are reported in the caption of Table \ref{tab:model} (all values are expressed in sec$^{-1}$). These values have been partly derived from literature \cite{Morton2}, and partly tuned to account for the following biological features \cite{Li,Mello}:
(1) the binding affinity of the ligand is directly proportional to the methylation state of MCPs;
(2) the ligand-receptor binding reactions occur at a faster rate with respect to phosphorylation and
methylation/demethylation reactions;
(3) the methylation and demethylation activities of CheR and CheBp are, respectively, inversely and directly proportional
to the methylation state of MCPs;
(4) the rate of phosphotransfer from CheA to CheY and CheB depends on the rate of autophosphorylation of CheA.
According to these constraints, which set the relative magnitude of some constants with respect to others, the estimation of the unavailable constants has been performed by testing the effect of a range of values
for each constant within every module of the model (that is, a group of reactions corresponding to a specific process, such as, e.g., reactions 1-4 that model the formation of the receptor complexes). Within this range, the finally chosen value for each constant is the one that gave a good reproduction of the expected behaviour of the biological subsystem described by that module. Then, every other module has been sequentially added to the previous ones, following the same iterative process, to perform a comprehensive and biologically correct simulation of the whole pathway.

\section{Stochastic simulations of the chemotactic response regulator}\label{sec:sim_dyn}

In this section, we briefly present the stochastic algorithm used to perform the simulation of the model defined in Section \ref{ssec:stochmod}. Then, we show the results of the dynamics of the chemotactic response regulator protein (the phosphorylated form of CheY) under different conditions of the chemotactic system.

\subsection{Tau leaping} \label{ssec:tau}

The algorithm called \emph{tau leaping} \cite{Gill06} is an approximated and faster version of the seminal Gillespie's stochastic simulation algorithm (SSA) \cite{SSA}. Both algorithms allow to generate the temporal evolution of chemicals contained inside a well stirred volume, in given and fixed experimental conditions. Chemicals interact with each other by means of given reactions, whose physical and chemical properties are encompassed in a specified stochastic constant associated to each reaction. Reactions are applied according to a probability distribution, that is determined -- at each computation step -- by the current state of the system (given by the number of molecules of each chemical) and by the value of all reaction constants. SSA and tau leaping share the characteristic that repeated (independent) executions will produce different temporal dynamics, even starting from the same initial configuration, thus reflecting the inherent noise of the system. The two algorithms differ with respect to the way reactions are applied. In SSA, only \emph{one} reaction can be applied at each step; the reaction that will be actually simulated, and the waiting time before the next reaction will take place, depend on two independent random numbers drawn from the uniform unit interval [0,1]. In tau leaping, instead, \emph{several} reactions can be chosen and executed simultaneously, by the sampling of Poissonian distributions and by choosing an opportune time increment (we refer to \cite{Gill06} for further details). So doing, the computational burden typical of SSA for large systems consisting of many reactions and many molecular species, can be largely reduced. Indeed, with tau leaping it is possible to handle detailed descriptions of many molecular interactions and chemical modifications (e.g., phosphorylation or methylation states of proteins), providing fast and reliable stochastic simulations of mechanistic models of complex biological systems \cite{cAMP}. On the other side, tau leaping is not guaranteed to reproduce the exact behavior of the system, but the accuracy of the simulation can be controlled.

Here we report a sketch of the functioning of tau leaping. We denote by $X$ a well stirred system in thermal equilibrium, consisting of $N$ molecular species $S_1, \dots, S_N$, which can interact through $M$ chemical reactions $r_1, \dots, r_M$. Let $X_i(t)$ be the number of molecules of chemical $S_i$ at time $t$, and $\mathbf{x} = \mathbf{X}(t) \equiv (X_1(t), \dots, X_N(t))$ the state of the system at time $t$. The aim of the procedure is to fire several reactions for each time interval $[t, t + \tau)$. In order to find out which reactions will be executed, we have to calculate the probability that a reaction $r_j$ will occur in the next infinitesimal time interval $[t, t+dt)$, starting from the system state $\mathbf{x}$. This probability is given by $a_j(\mathbf{x})dt$, which is the \emph{propensity function} of reaction $r_j$ and is defined as $a_j(\mathbf{x}) = h_j(\mathbf{x}) \cdot c_j$, where $h_j(\mathbf{x})$ is the number of distinct reactant molecules combinations in $r_j$, and $c_j$ is the stochastic  constant associated to $r_j$.

Given a state $\mathbf{x}$ of the system $X$, we denote by $K_j(\tau, \mathbf{x}, t)$ the exact number of times that a reaction $r_j$ will be fired in the time interval $[t, t + \tau)$, so that $K(\tau, \mathbf{x}, t)$ is the exact probability distribution vector (having $K_j(\tau, \mathbf{x}, t)$ as elements). For arbitrary values of $\tau$, the computation of the values of $K_j(\tau, \mathbf{x}, t)$ is as difficult as resolving the corresponding Chemical Master Equation for that system. On the contrary, if $\tau$ is small enough so that the change in the state $\mathbf{x}$ during $[t, t + \tau)$ is so slight that no propensity function will suffer an appreciable change in its value (this is called the \emph{leap condition}), then it is possible to evaluate a good approximation of $K_j(\tau, \mathbf{x}, t)$ by using the Poisson random variables with mean and variance $a_j(\mathbf{x})\cdot\tau$. Hence, starting from the state $\mathbf{x}$ and choosing a $\tau$ value that satisfies the leap condition, we can update the state of the system at time $t + \tau$ according to
$\mathbf{X}(t+\tau) = \mathbf{x} + \sum_{j=i, \dots, M} \mathbf{v}_j P_j(a_j(\mathbf{x}),\tau),$
where $P_j(a_j(\mathbf{x}), \tau)$ denotes an independent sample of the Poisson random variable with mean and variance $a_j(\mathbf{x})\cdot\tau$, and $\mathbf{v}_j \equiv (v_{1j}, \dots, v_{Nj})$ is the state change vector whose element $v_{ij}$ represents the stoichiometric change of the species $S_i$ due to reaction $r_j$. Summarizing, each iterative step of the algorithm consists of four stages: (1) generate the maximum changes of each species that satisfy the leap condition; (2) compute the mean and variance of the changes of the propensity functions; (3) evaluate the leap value $\tau$ exploiting the auxiliary quantities previously computed; (4) toss the reactions to apply; (5) update the system state (see \cite{Gill06} for further details).

The accuracy of this algorithm can be fixed a priori by means of an error control parameter $\epsilon$ ($0<\epsilon \leq 1$); for the following simulations, we have used the value $\epsilon = 0.03$ as suggested in \cite{Gill06}. The algorithm has been implemented in our laboratory \cite{tauDPP}; the program is written in C language,
compiled under Linux using the GCC compiler. All the simulations have been performed using a Personal Computer with an
Intel Core2 CPU (2.66 GHz) running Linux. The mean duration time for one run, for the simulation of the dynamics of CheYp over 3000 sec, is about 4-5 seconds (with the initial values of chemical amounts given in Section \ref{ssec:stochmod} and the constants reported in Table \ref{tab:model}). All the figures reported in Section \ref{ssec:sim_results}, unless otherwise stated, represent the mean value over 50 independent runs of tau leaping, each one executed with the same initial conditions.

\subsection{Results of simulations for the dynamics of CheYp} \label{ssec:sim_results}

The dynamics of CheYp has been analyzed by considering various conditions, such as the addition and removal of different ligand amounts, distinct methylation states of MCPs and deletion of other chemotactic proteins.

\begin{figure}[htpb]
\centerline{\includegraphics[width=4.5cm,angle=-90]{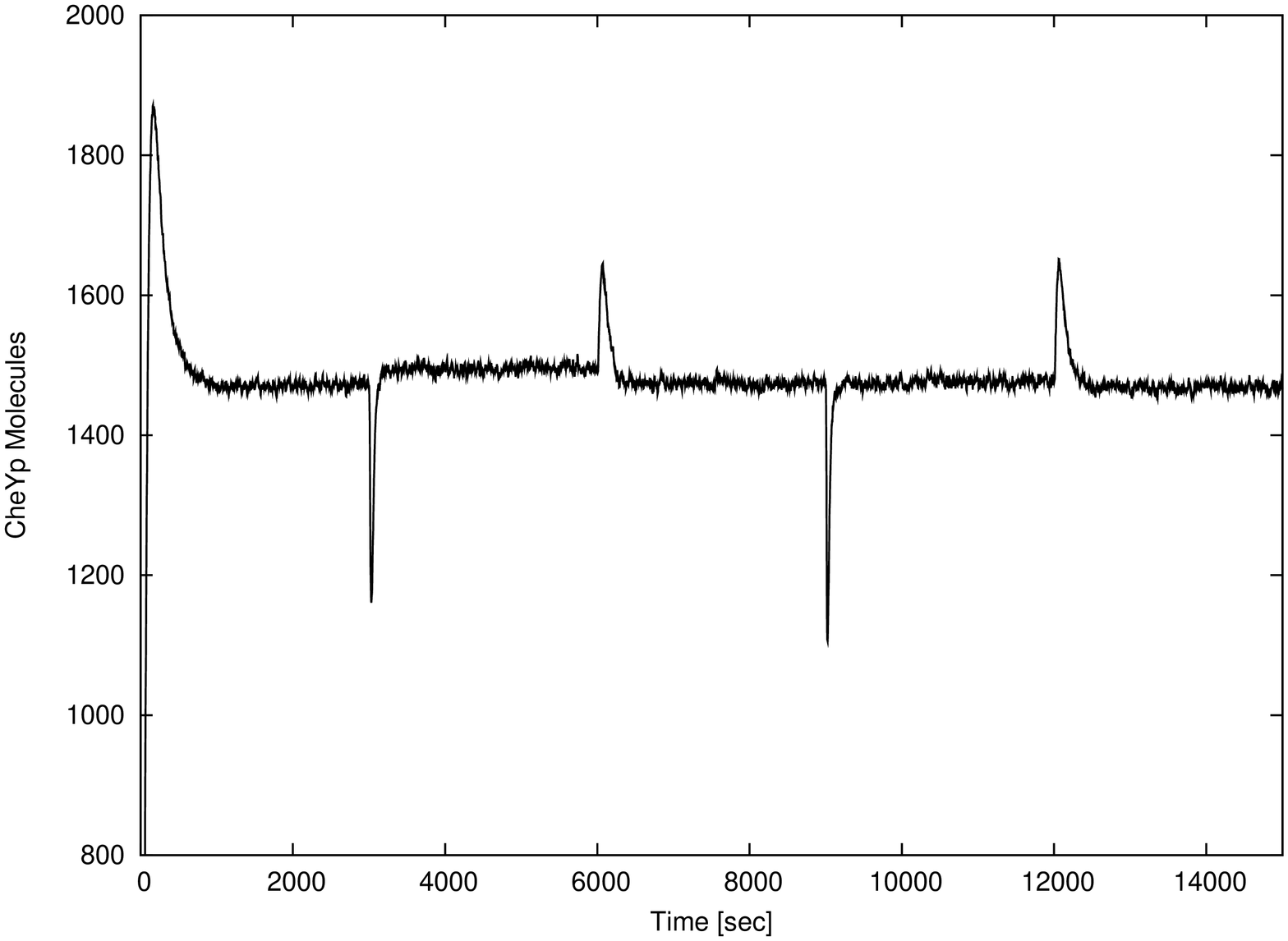} 
\includegraphics[width=4.5cm,angle=-90]{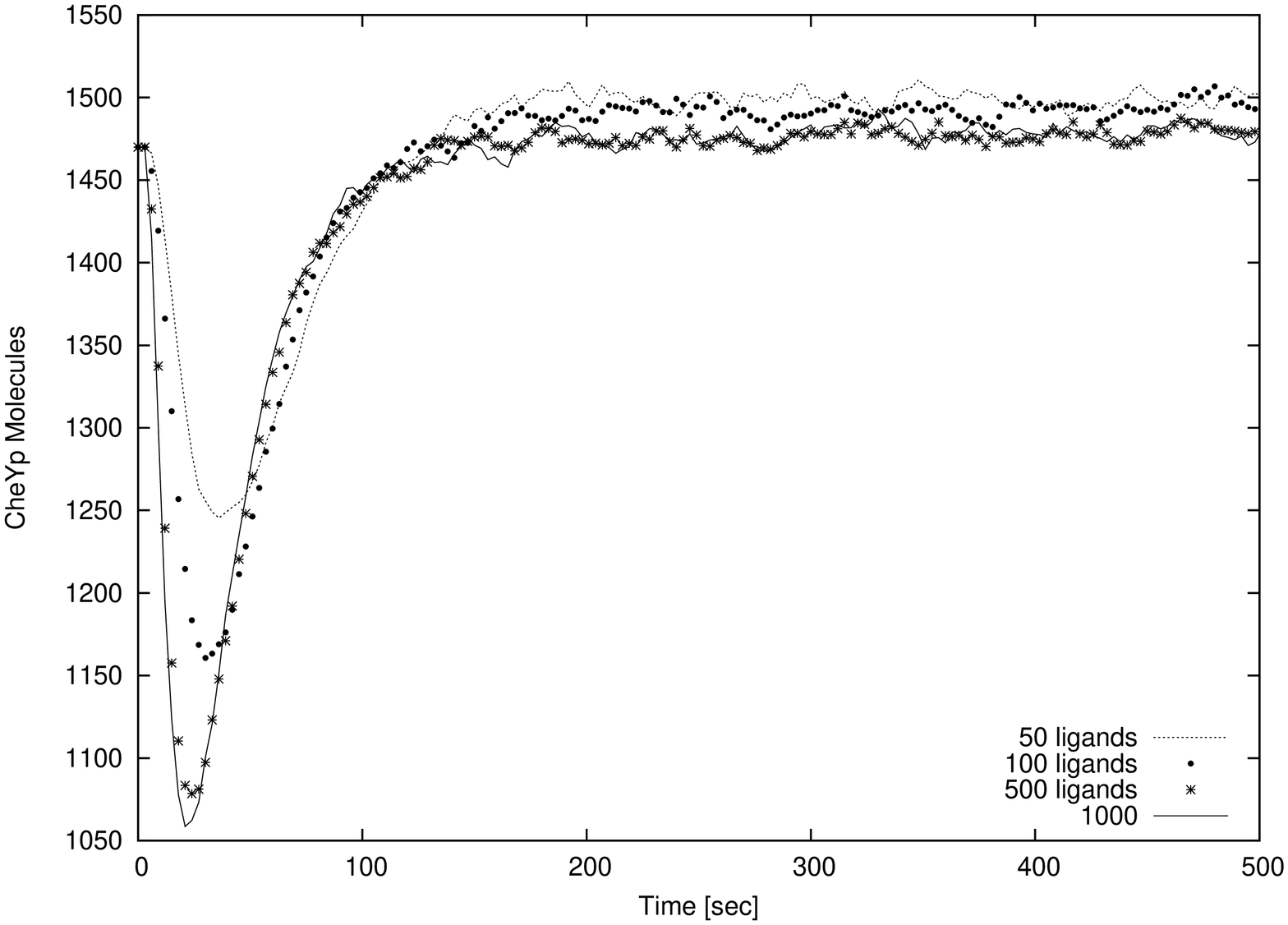}} \caption{Dynamics of CheYp. Left:
adaptation response to two consecutive stimuli. Right: comparison of transient and steady state response to
different ligand amounts.} \label{fig:succ_stimuli}
\end{figure}

We start by reporting in Figure \ref{fig:succ_stimuli}, left side, the response of the system to the addition of two consecutive amounts of external ligand: the first stimulus corresponds to a ligand amount of 100 molecules, added to the system at time $t=3000$ sec and removed at time $t=6000$ sec, while the second stimulus corresponds to a ligand amount of 500 molecules, added at time $t=9000$ sec and removed at time $t=12000$ sec. Note that, since the amount of CheYp is equal to 0 at the beginning of the simulation, its dynamics shows a marked increase which then reaches -- due to the counteraction of CheZ, CheR and CheB -- a steady state level. Starting from this level, the addition of the ligands has been simulated by changing its amount from 0 to 100 (500, respectively) molecules, thus mimicking the environmental situation where the bacterium encounters a different concentration of attractant molecules. Vice versa, the removal of the ligands has been simulated by putting the value of the ligand back to 0. In the time interval between the addition and the removal of each ligand stimulus, the amount of ligand molecules has been kept at the constant value of 100 and 500, respectively, thus mimicking the presence of an environmental homogenous concentration. This has been done in order to test the adaptation capabilities of the system. In both cases, we can see that the system is able to respond to a step-increase of the ligands by achieving a sharp and fast decrease in CheYp (that is, the negative peaks at time instants $t=3000$ and $t=9000$ sec). Immediately after this transient, the amount of CheYp returns to a steady state value, which differs from the prestimulus level only for a few tens of molecules, at most, according to the amount of added ligand. In this phase, the bacterium is returning to the prestimulus switching and thus to the random walk sampling of its surroundings. When the ligand is removed, CheYp shows another transient
behavior, corresponding to a sharp and fast increase of its amount, that is in line with experimental observations (see \cite{BarkaiLeibler,RaoPLOS}). After this second transient, the amount of CheYp correctly returns to the prestimulus steady state level.

\begin{table}\caption{Steady state values and minimal/maximal transient values of CheYp after addition and removal
of distinct ligand amounts.}\label{tab:steadystates}
\begin{center}
\begin{small}
\begin{tabular}{|l|l|l|l|l|l|}
  \hline
 \textbf{Ligand amount} & \textbf{SS1} & \textbf{Min} & \textbf{SS2} & \textbf{Max}& \textbf{SS3}\\
  \hline
50 molecules & 1486.7 & 1245.4 & 1500.9 & 1626.0 & 1474.7\\
100 molecules & 1486.7 & 1160.7 & 1495.1 & 1645.4 & 1474.3\\
500 molecules & 1486.7 & 1078.4 & 1481.4 & 1653.2 & 1469.4\\
1000 molecules & 1486.7 & 1058.6 & 1478.2 & 1665.8 & 1474.7\\
  \hline
\end{tabular}
\end{small}
\end{center}
\end{table}

In Figure \ref{fig:succ_stimuli}, right side, we compare the transients and steady states reached by CheYp
after the addition of distinct ligand amounts. This figure shows that the response magnitude at steady state and the adaptation time of CheYp is only slightly sensitive to the ligand amount, being the relative differences less than a few tens of molecules and less than a few seconds, respectively. The mean values of the steady state of CheYp before the stimulus (SS1), after the ligand addition (SS2) and after the ligand removal (SS3) are reported in Table \ref{tab:steadystates}, together with the values of its minimal and maximal values immediately after the ligand addition and removal (Min and Max, respectively), for the four ligand amounts (50, 100, 500, 1000 molecules) considered in the right side of Figure \ref{fig:succ_stimuli}.

\begin{figure}[htpb]
\centerline{\includegraphics[width=4.5cm,angle=-90]{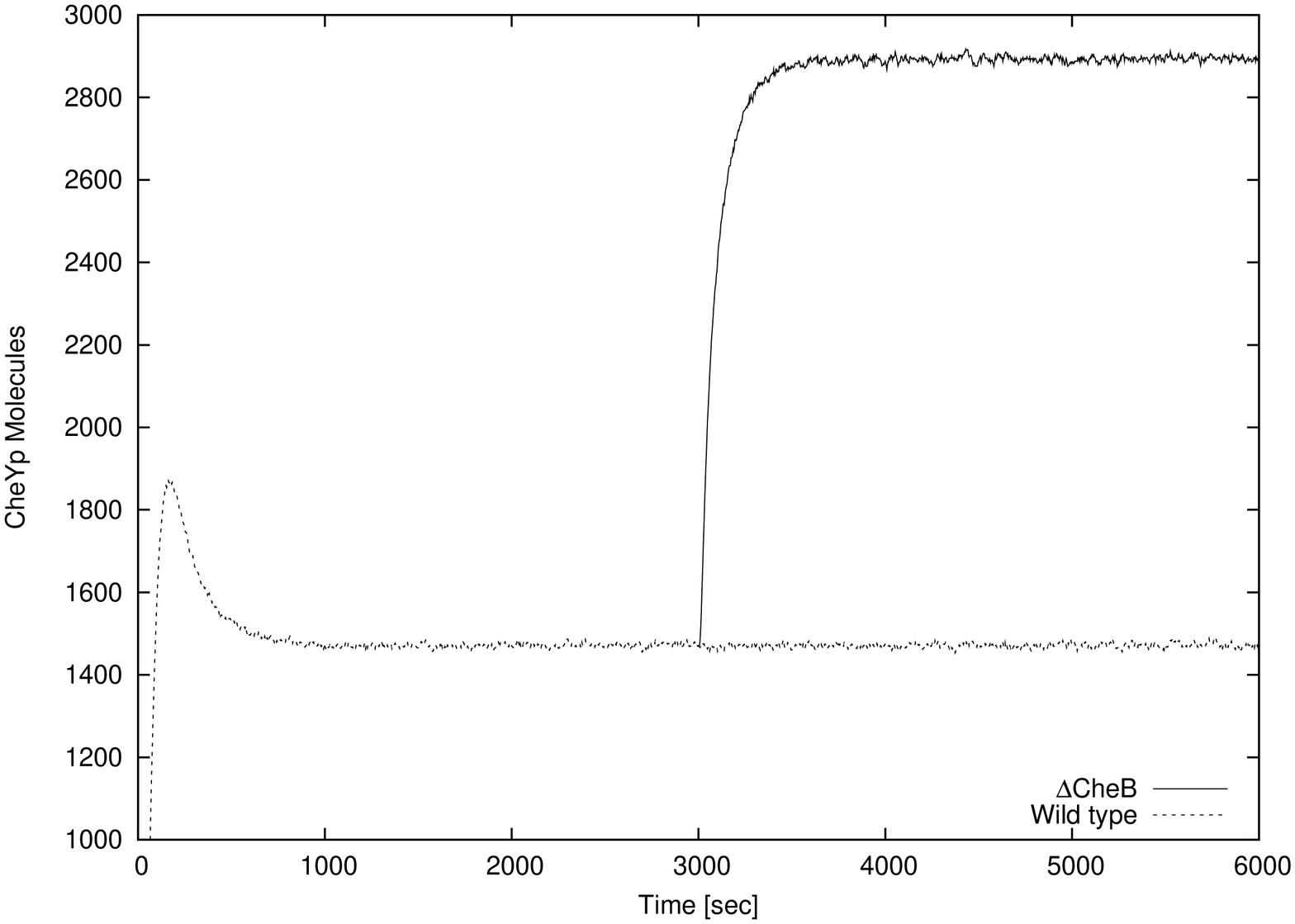} 
\includegraphics[width=4.5cm,angle=-90]{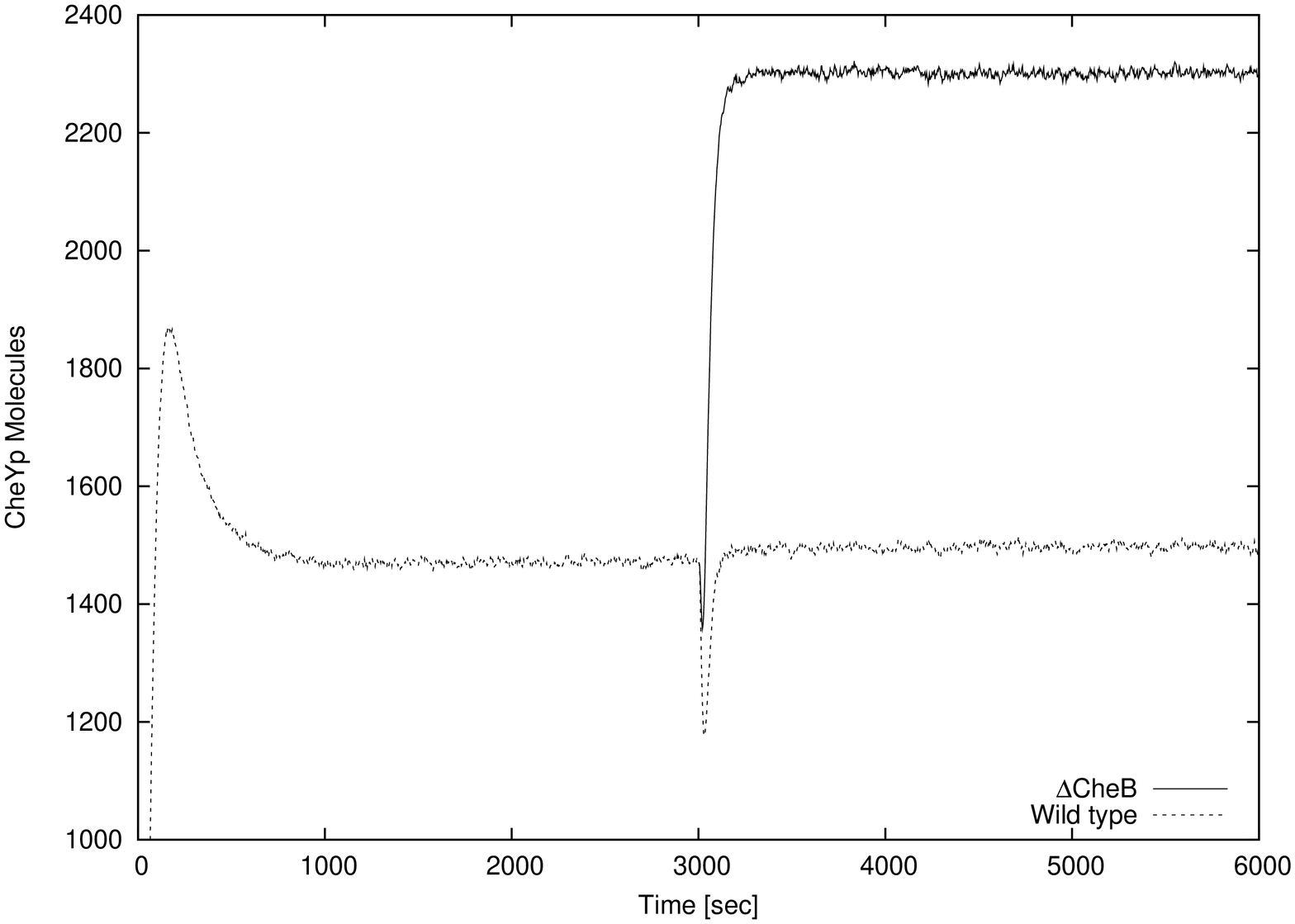}} \caption{Comparison of dynamics of CheYp in normal condition
and after deletion of CheB at $t=3000$ sec, without ligand (left) and with simultaneous addition of 100 ligand molecules
(right).} \label{fig:noCheB}
\end{figure}

In Figure \ref{fig:noCheB} we show how the dynamics of CheYp changes when CheB is deleted from the system at time $t=3000$ sec, in both conditions of absence of external ligands (left side) and of presence of 100 molecules of ligand (right side) added at time $t=3000$ sec.
CheB is the methylesterase that, once being phosphorylated by CheA, increases the methylation state of MCPs, thus keeping CheA more active.
This, in turn, causes an increase in the amount of CheYp, which is evident from its new steady state level reached after CheB deletion, and also from its less negative transient decrease after ligand addition.

\begin{figure}[htpb]
\centerline{\includegraphics[width=4.5cm,angle=-90]{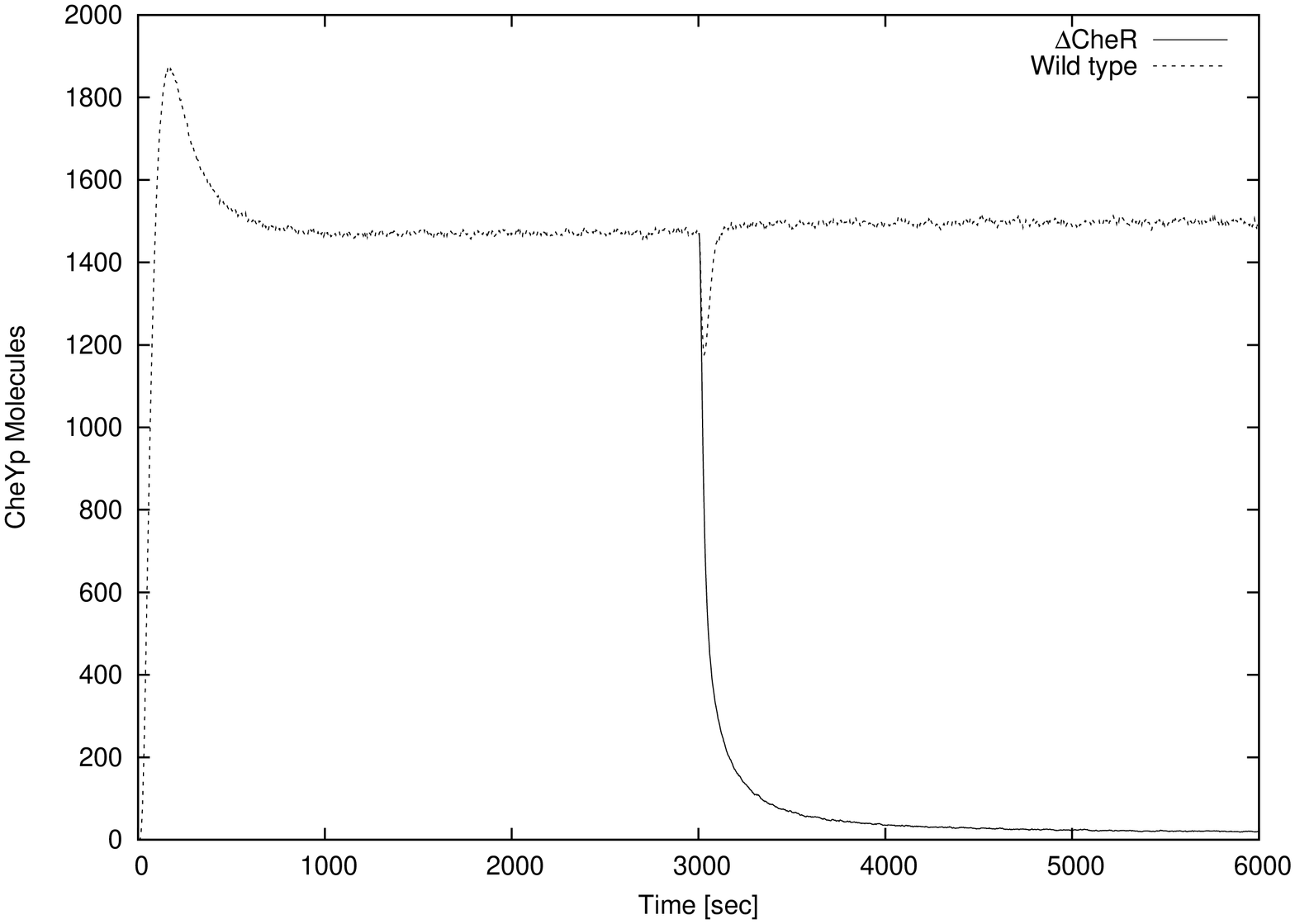} 
\includegraphics[width=4.5cm,angle=-90]{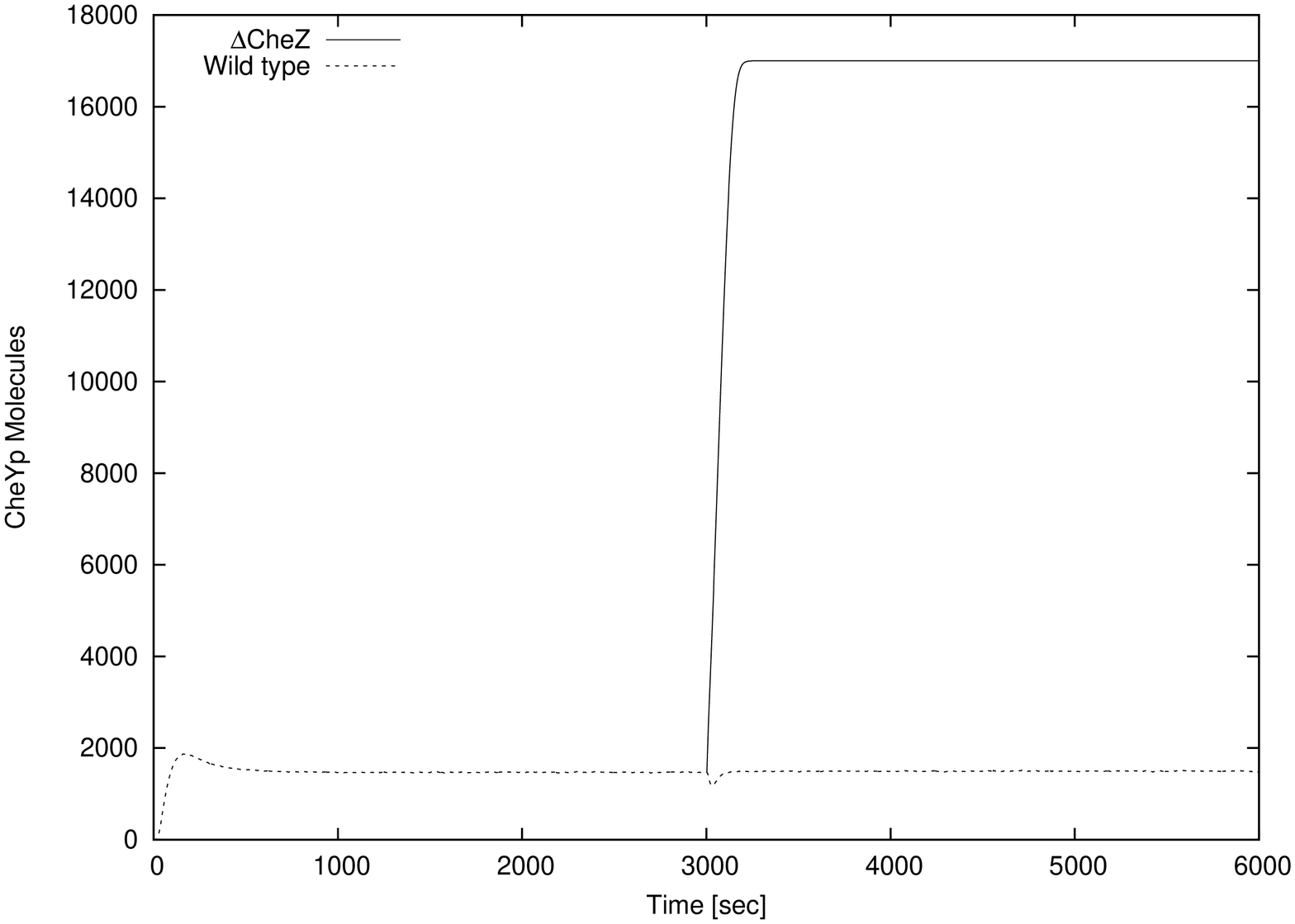}} \caption{Comparison of dynamics of CheYp in normal condition and
after deletion of CheR (left) and CheZ (right) at $t=3000$ sec, both simulated with a simultaneous addition of 100 ligand molecules.} \label{fig:noCheR-Z}
\end{figure}

\begin{figure}[htpb]
\centerline{\includegraphics[height=7cm,angle=-90]{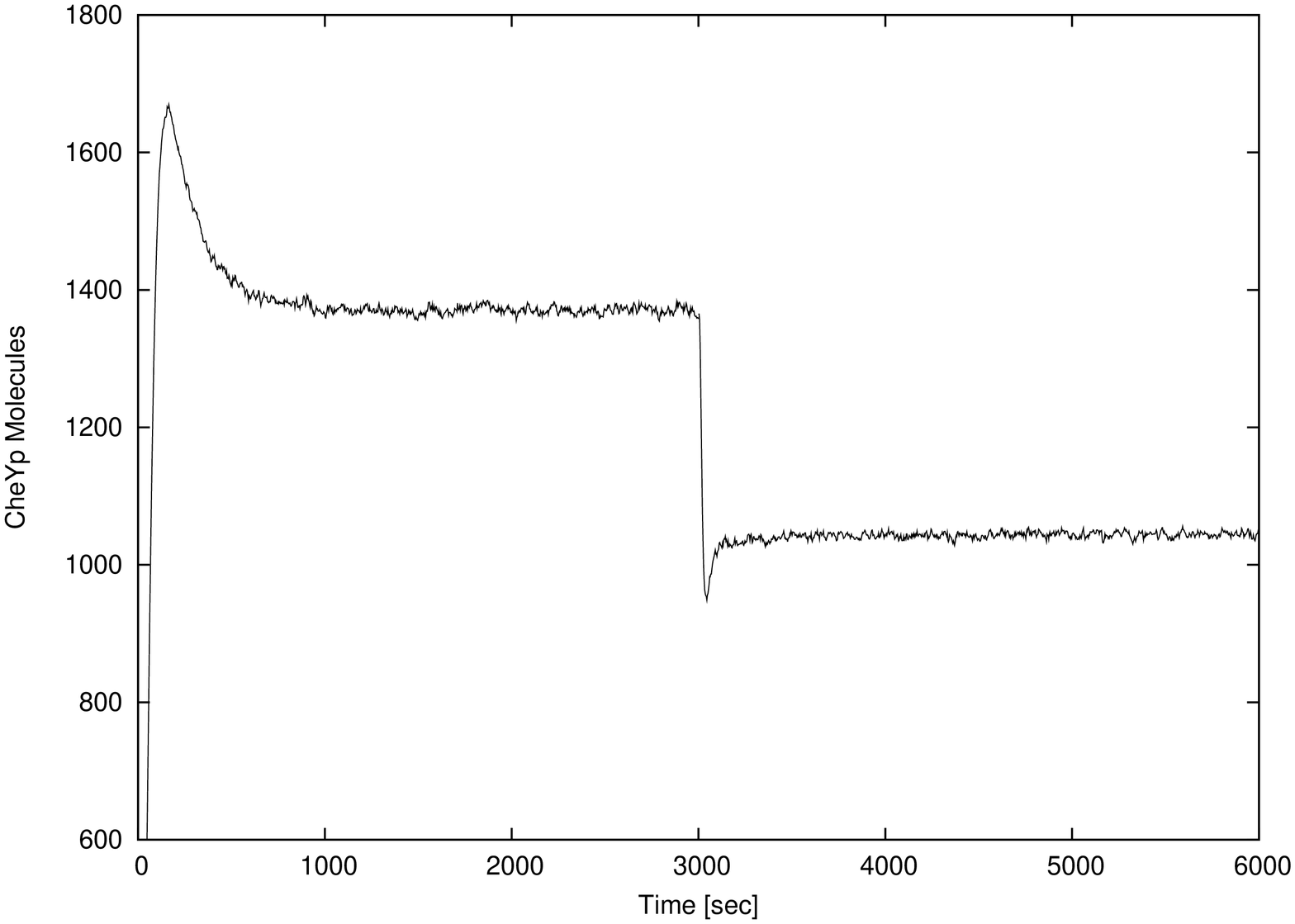}\includegraphics[height=7cm,angle=-90]{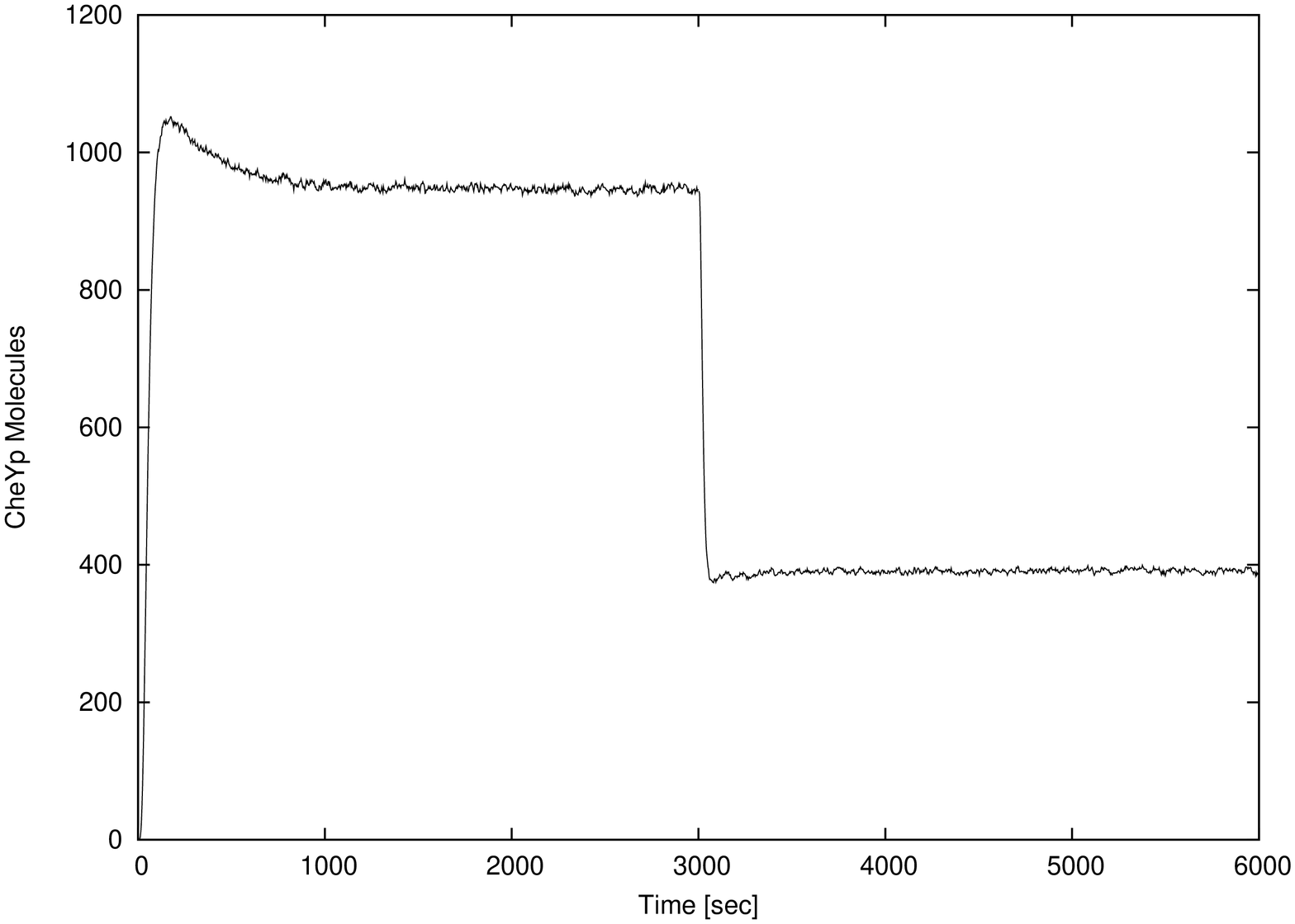}}
\caption{Dynamics of CheYp when only 3 (left) and 2 (right) methylation states are active.} \label{fig:metlevels}
\end{figure}

Similarly, in Figure \ref{fig:noCheR-Z} we show the dynamics of CheYp when either CheR (left side) or CheZ (right side) are deleted from the system at time $t=3000$ sec, simultaneously to the addition of 100 ligand molecules (the temporal evolution of CheYp when no ligand is added is basically equivalent). When CheR is deleted, its methyltransferase activity is silenced, the MCPs are no more methylated, and hence the amount of CheYp tends to zero. On the contrary, when CheZ is deleted, all CheY molecules always remain phosphorylated. For the sake of completeness, we have also simulated the dynamics of CheYp when either CheB, CheR or CheZ are deleted from the system since the time instant $t=0$, in order to have a comparison about the initial temporal evolution of CheYp
and the steady state levels it can reach. In these conditions, the model correctly simulates \cite{robust,Kim,Scharf} a very low production of CheYp when CheR is deleted, and an increased production (albeit with different magnitudes) when either CheB or CheZ are deleted (data not shown).

Finally, in Figure \ref{fig:metlevels} we compare the dynamics of CheYp in response to the addition of 100 ligand molecules at $t=3000$ sec, when only 3 (left side) or 2 (right side) methylation states of the receptors are allowed. In practice, this is achieved by initially putting to zero the values of the stochastic constants of methylation and demethylation reactions for levels $m=4$ and $m=3$, respectively. In both cases, we see that the system
is not able to adapt, as the steady state level of CheYp reached after the addition of the ligand is substantially lower than the steady state when all methylation levels are activated.

The outcome of the stochastic simulations reported in this section validates the model presented in Section \ref{ssec:stochmod}, as the dynamics of CheYp under different conditions of the chemotactic pathway well compare with experimental evidences.

\section{The interplay between stochastic fluctuations and the number of bacterial flagella}\label{sec:interplay}

In this section we make use of the simulations based on our model of chemotaxis to analyze the interplay between stochastic fluctuations of CheYp and the number of flagella occurring on the cell, in order to outline the influence of synchronization of flagellar motors on the swimming behavior, and on the adaptation mechanism of the bacterium to the environmental changes. To this aim, we consider the dynamics of CheYp at steady state, as well as its transient step-decrease that takes place immediately after the chemotactic stimulus. In both cases, we are interested in devising the time periods during which the cell performs either a running or a tumbling motion. In particular we will assume that: (1) the time spent in alternating CW and CCW rotations during the steady state corresponds to the random walk sampling of the environment -- where we expect more time spent in tumbling than in running motions; (2) the time required to return to the prestimulus level of CheYp (that is, the transient response immediately after the ligand addition) corresponds to the chemotactic adaptation time -- where we expect a much longer and uninterrupted time interval of running motion with respect to the steady state condition.

As explained in Section \ref{ssec:bio}, a running motion requires that all flagella are simultaneously synchronized in a CCW rotation -- which occurs when CheYp is not interacting with the proteins FliM of the flagellar motors, that is, when its intracellular concentration diminishes with respect to a reference value. To distinguish between the CW and CCW rotations of a \emph{single} flagellum, we assume that the flagellar motor switch is sensitive to a threshold level of CheYp, that is hereby evaluated as the mean value of CheYp at steady state (see also \cite{Morton}, where a similar approach of threshold-crossing mechanism for motor switching was tested, albeit that work considered only a single flagellum and did not propose any investigation on the simultaneous coordination of many flagella). When the amount of CheYp is below this threshold, each flagellum is rotating CCW, while when the amount of CheYp is above the threshold, each flagellum is rotating CW. In what follows, we make a one-to-one correspondence between the behavior of a single flagellum and a temporal evolution of CheYp generated by one run of the tau leaping algorithm, that is, we consider a different and independent stochastic simulation for each and every flagellum (albeit starting from the same initial conditions for the whole system). In other words, we assume that flagella are independent one another -- as no molecular interactions between them have been evidenced in bacterial cells. Nonetheless, they all overlook on the same intracellular ambient, that is, they are all subject to the same temporal evolution of CheYp, apart from the stochastic noise existing among independent simulations.
In order to determine the synchronization of all flagella that will induce a running motion of the bacterium, we therefore need to identify the time instants when \emph{all} flagella are rotating CCW, that is, to select the time intervals when \emph{all} the temporal evolutions of CheYp are below the fixed threshold.

\begin{figure}
\centerline{\parbox{7.5cm}{\includegraphics[width=6.5cm]{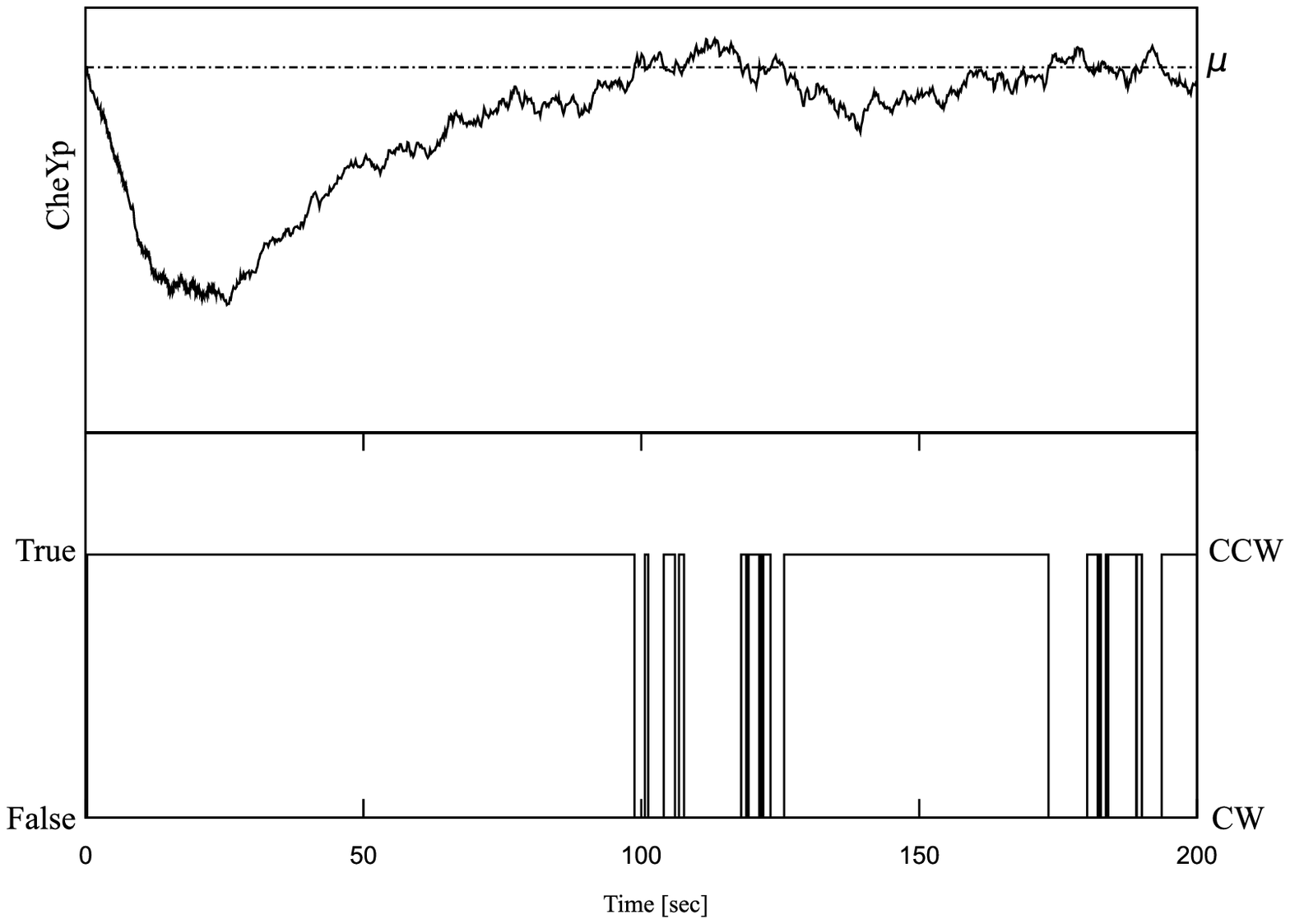}}
\parbox{7.5cm}{\includegraphics[width=6.5cm]{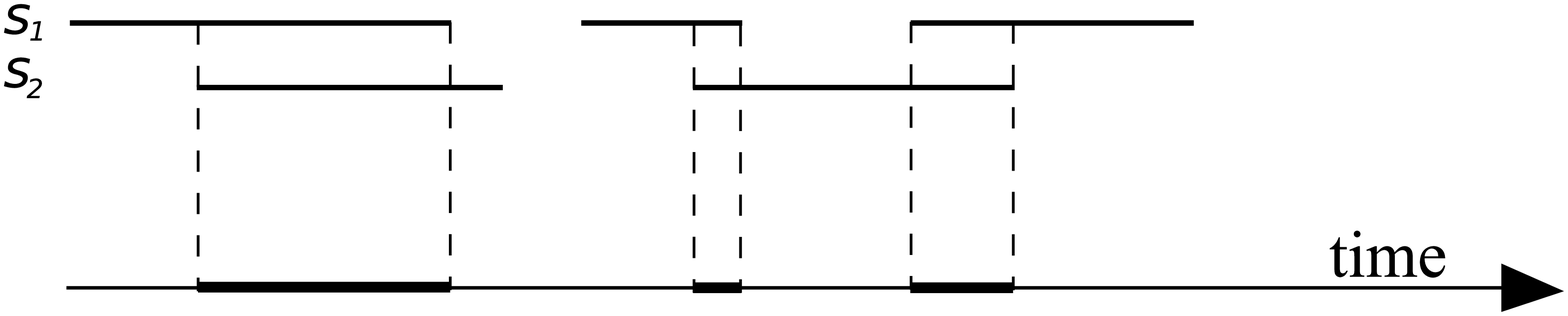}}} \caption{Threshold-crossing intervals in stochastic fluctuations of CheYp (left) and synchronization of running motion between 2 flagella (right).} \label{fig:stoch_fluct_true}
\end{figure}

Formally, we proceed as follows. Let $n=1, \dots, 10$ be the number of flagella $f_1, \dots, f_n$ whose influence we want to test, and let $s_i$, $i=1, \dots, n$, be the time series of CheYp (generated by a single tau leaping run) associated to each $f_i$. For any fixed value of $n$, the total time of the simulation considered to generate the dynamics of CheYp is the same for all $s_i$. This simulation time, hereby denoted by $\Delta t_{sim}$, is chosen long enough to have a meaningful evaluation of the mean intervals of running and tumbling in the analysis performed below (e.g., $\Delta t_{sim}=40000, 60000, 120000$ sec for $n=1,5,10$, respectively). The threshold for CheYp is evaluated in the following way: we choose an initial time instant at the steady state level -- distant enough from the step decrease of CheYp after ligand addition, i.e. 1000 sec afterward -- and then, starting from this instant and till the end of $\Delta t_{sim}$, we calculate the mean value $\mu_i=<s_i>$ for each $s_i$. Then, we define a common threshold $\mu$ for all flagella, that is, $\mu=\frac{1}{n} \sum_{i=1, \dots, n} \mu_i$. This threshold is then considered as the reference value also for the portion of the CheYp dynamics corresponding to the transient decrease after ligand addition. In Figure \ref{fig:stoch_fluct_true}, top panel on the left side, we show a part of $\Delta t_{sim}$ over a single simulation of CheYp, where both the initial transient response and the stochastic fluctuations around the threshold are evidenced. For all the results discussed below, the different values of $\mu$ have been found to be approximatively equal to 1480 molecules.

The next step consists in detecting, for each $f_i$, the time intervals during which the amount of CheYp remains below $\mu$, each one of these intervals corresponding to a CCW rotation time interval of that flagellum. Namely, for each $s_i$ we identify the time intervals $\Delta t_{true} \subseteq \Delta t_{sim}$ such that $\Delta t_{true}=\{t\in \Delta t_{sim} \mid s_i(t) - \mu \leq 0\}$. Note that this simple mechanism of single threshold-crossing could be extended to consider more complex situations -- e.g., a double threshold-crossing mode can be assumed -- whereby one simply asks for analogous conditions to be satisfied. Similarly, for each $s_i$ we can locate the complementary time intervals $\Delta t_{false} \subseteq \Delta t_{sim}$ such that $\Delta t_{false}=\{t\in \Delta t_{sim} \mid s_i(t) - \mu > 0\}$; these intervals correspond to the time that each flagellum $f_i$ spends in a CW rotation. Stated in other terms, we can associate to each $s_i$ a function $CCW_{s_i}: \Delta t_{sim} \rightarrow \{true,false\}$ defined as:
$$CCW_{s_i}(t)=\left\{
    \begin{array}{ll}
    true & \mbox{if } s_i(t) - \mu \leq 0 \\
    false & \mbox{otherwise}
    \end{array}
\right.$$

In Figure \ref{fig:stoch_fluct_true}, bottom panel on the left side, we show the values of this function for the CheYp simulation given in the upper panel. As it can be seen at a glance, the transient response after ligand addition (when the amount of CheYp is initially below $\mu$) corresponds to a longer and uninterrupted interval of CCW rotation of that flagellum.

Once that the set of all $\Delta t_{true}$ intervals -- or, equivalently, of all functions $CCW_{s_i}$ -- have been found out for each flagellum, we start the process of synchronization for any given number $n$ of flagella. To this aim, let us define $\mathcal{T}^n_{sync}=\{t \in \Delta t_{sim} \mid CCW_{s_i}(t)=true \mbox{ for all } i=1, \dots, n\}$. $\mathcal{T}^n_{sync}$ is the set of all times during which \emph{all} time series $s_i$ are below the threshold $\mu$, that is, the time intervals during which \emph{all} flagella are rotating CCW. More precisely, we identify these intervals as the running motion of the bacterium, i.e. $\mathcal{T}^n_{sync}$ corresponds to the time of directional swimming -- when all flagella are coordinated in a bundle. As an example, in Figure \ref{fig:stoch_fluct_true}, right side, we represent the functions $CCW_{s_i}(t)=true$ for $i=1,2$, and the corresponding set $\mathcal{T}^n_{sync}$, $n=2$. The complementary set, $\mathcal{T}^n_{unsync}=\Delta t_{sim} \setminus \mathcal{T}^n_{sync}$, corresponds instead to tumbling motion -- when at least one flagellum (over the set of $n$ flagella considered time by time) is rotating CW. Namely, $\mathcal{T}^n_{unsync}=\{t \in \Delta t_{sim} \mid \mbox{ there exists } i=1, \dots, n \mbox{ such that } CCW_{s_i}(t)=false\}$.

We are now interested in understanding if and how the time intervals within the set $\mathcal{T}^n_{sync}$ are influenced by the increase of $n$. We have performed this analysis over a set of 10 distinct in silico experiments (each one corresponding to a cell with $n$ flagella, with $n=1, \dots, 10$), and then we have evaluated the mean values of the following three parameters:
\begin{enumerate}
\item the time intervals corresponding to a running motion of the bacterium, $\langle \Delta t_{run} \rangle$, when all flagella are rotating CCW (that is, when all time series $s_i$ are below $\mu$);
\item the time intervals corresponding to a tumbling motion  of the bacterium, $\langle \Delta t_{tumb} \rangle$, when at least one flagellum over the $n$ flagella is rotating CW (that is, when at least one time series $s_i$ is above $\mu$);
\item the time intervals corresponding to the transient decrease of CheYp after ligand addition, $\langle \Delta t_{adapt} \rangle$, that is, the adaptation time during which the bacterium is performing a longer running motion.
\end{enumerate}

\begin{figure}
\centerline{\includegraphics[height=3.8cm]{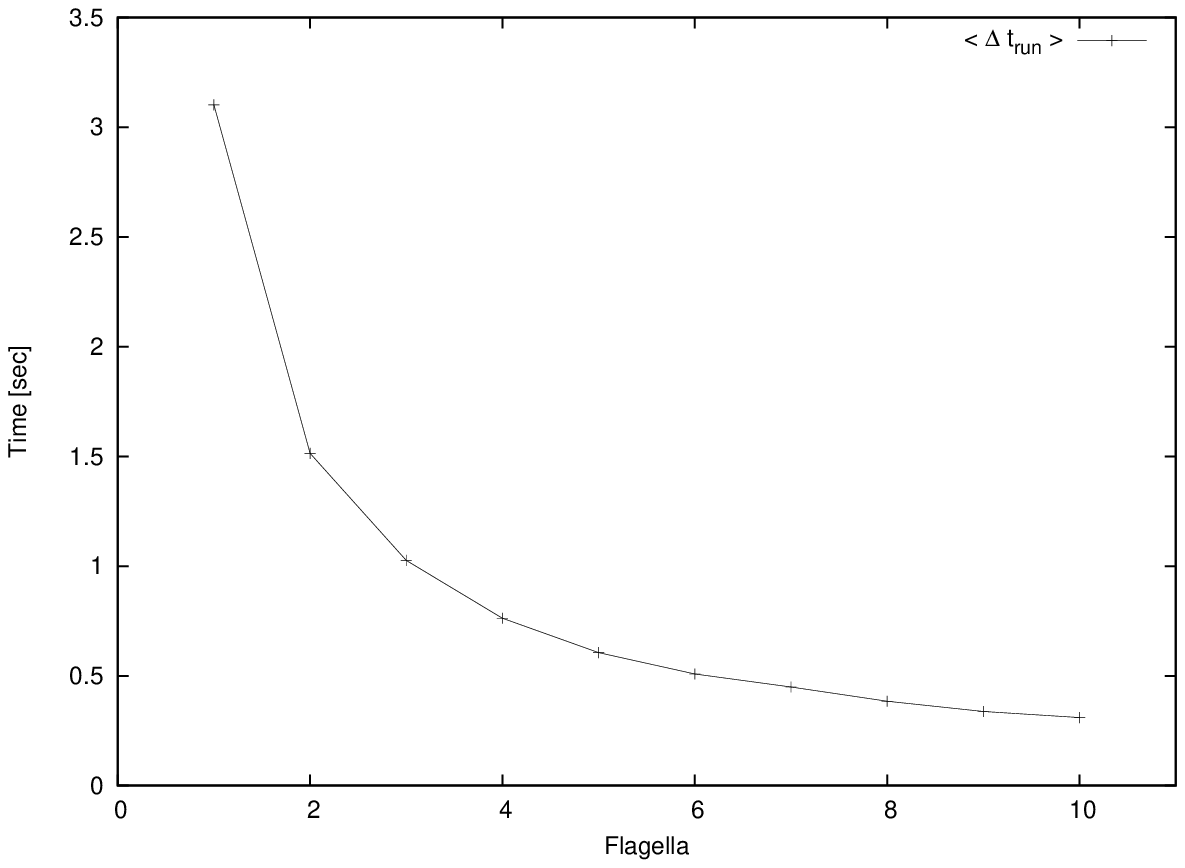}
\includegraphics[height=3.8cm]{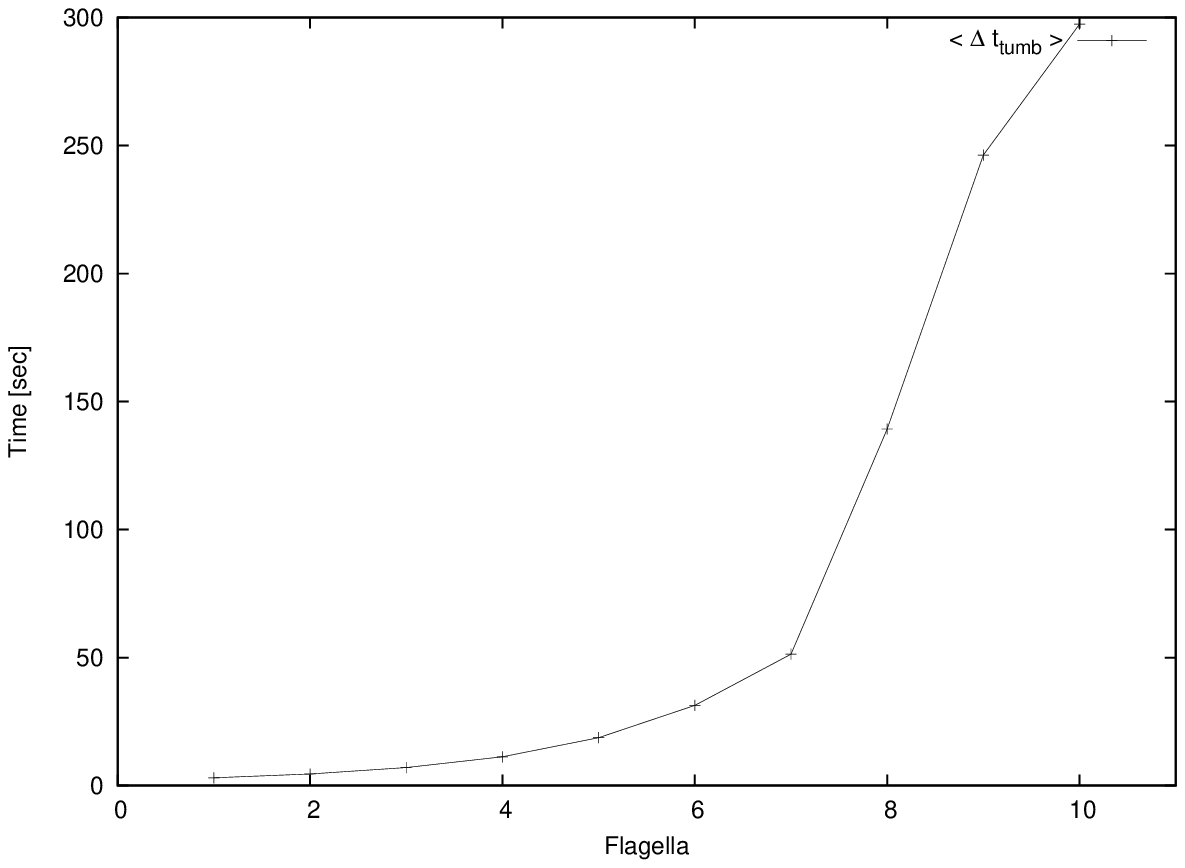}
\includegraphics[height=3.8cm]{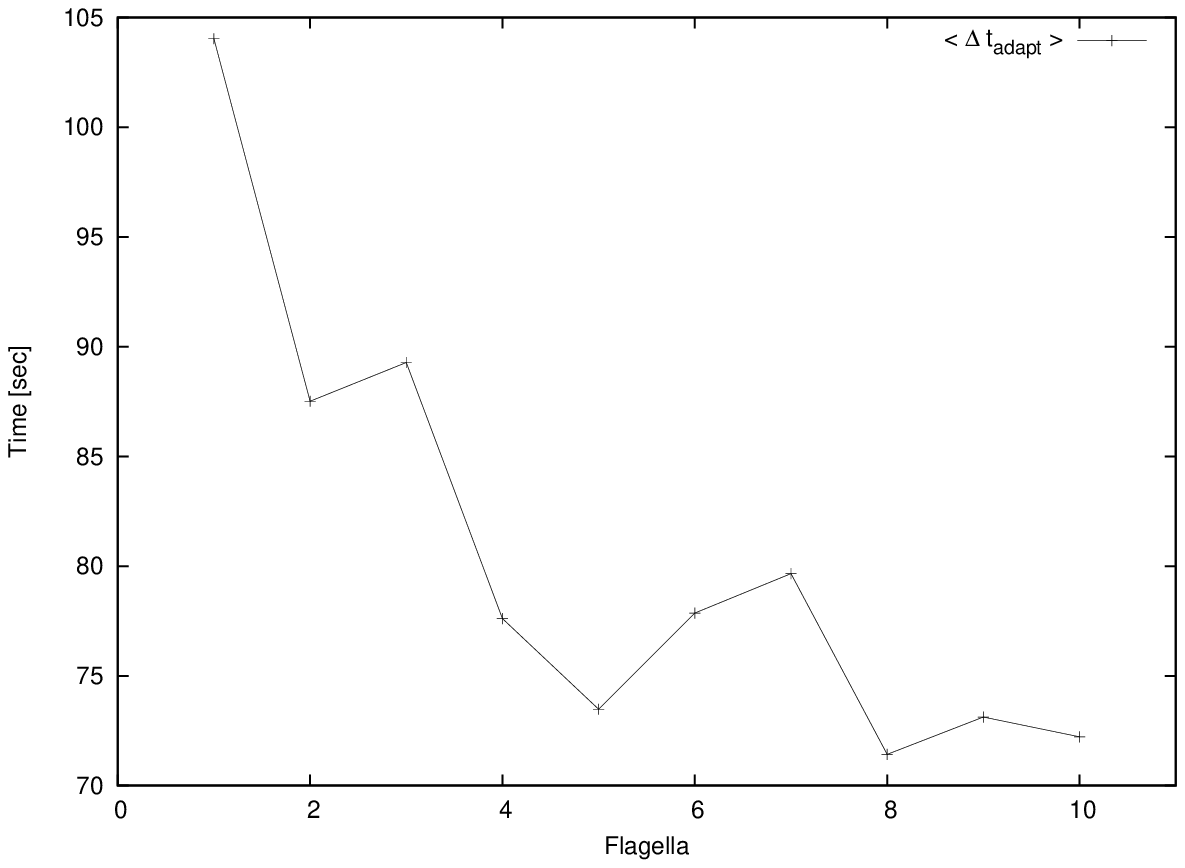}}
\caption{Variation of mean time values of running (left) and tumbling motions (middle), and of adaptation time (right), with respect to the number of flagella.}
\label{fig:mean_interv}
\end{figure}

\begin{table}\caption{Values of mean time intervals for running, tumbling and adaptation.}\label{tab:mean_values}
\begin{center}
\begin{small}
\begin{tabular}{|c|c|c|c|c|}
\hline
n & $\langle \Delta t_{run} \rangle$ (sec) &  $\langle \Delta t_{tumb} \rangle$ (sec) & $\langle \Delta t_{run} \rangle / \langle \Delta t_{tumb} \rangle$ & $\langle \Delta t_{adapt} \rangle$ (sec)\\
\hline
1 & 	3.102	& 3.062	& 1.013 & 104.0 \\
5 &	    0.606   & 18.73 & 0.032 & 73.48 \\
10 &	0.310	& 297.4	& 0.001 & 72.22 \\
\hline
\end{tabular}
\end{small}
\end{center}
\end{table}

The results for $\langle \Delta t_{run} \rangle$ are reported in Figure \ref{fig:mean_interv}, left side, where we can see that the mean time intervals of running motion are very short, and their values decrease in a (qualitative) exponential way as the number $n$ of flagella increases, as expected. Similarly, the results for $\langle \Delta t_{tumb} \rangle$  evidence a (qualitative) exponential increase with respect to $n$, as reported in Figure \ref{fig:mean_interv}, middle part. As reference, the precise values of the mean running and tumbling time intervals are given in Table \ref{tab:mean_values}, together with their ratio, for three values of $n$. The running-to-tumbling ratio, which decreases as $n$ increases, highlights the relevance of the number of flagella and the necessity of their synchronization with respect to the chemotactic behavior of the bacterium. That is, we see that for $n=1$ the time spent in running or tumbling motions is approximatively equivalent, but if coordination among many flagella ($n=10$) has to take place, then the running motions are highly reduced with respect to tumbling motions, which is in agreement with biological expectations.

The results for $\langle \Delta t_{adapt} \rangle$ are reported in Figure \ref{fig:mean_interv}, right side, and in Table \ref{tab:mean_values}. In this case, it is not possible to recognize a simple function for the curve progress, and we see that the variation of the time intervals is within a range of a few tens of seconds. Once more, this result seems to be in agreement with biological expectations, as the response of the bacterium to an environmental change (i.e. the addition or removal of ligands) should not be strictly dependent on the number of flagella that are present on its surface, otherwise the chemotactic pathway would not guarantee an appropriate adaptation mechanism, independently from the variation of the number of flagella among distinct individuals in a population of cells.

\section{Discussion}\label{concl}

In this paper we have investigated the possible influence of stochastic fluctuations of the chemotactic protein CheYp on the running motion of bacterial cells, with respect to an increasing number of flagella in the individual bacterium. To this aim, we have defined a procedure to identify the synchronization of CCW rotations of each and every flagella, and then we have compared the mean time intervals of running and tumbling motions of the cell, as well as of adaptation times to ligand addition, according to the different numbers of flagella. We have shown that, on the one hand, the running-to-tumbling ratio highlights the relevance of the number of flagella, and the necessity of their synchronization with respect to the chemotactic behavior of the bacterium. On the other hand, the adaptation time does not seem to be strongly influenced by the varying number of flagella in distinct individual cells. These results have been obtained by performing stochastic simulations of a very detailed mechanistic model of the bacterial chemotaxis pathway, that takes into account all proteins, and their respective interactions, involved in both signaling and response. All post-translational modifications of proteins, such as methylation and phosphorylation, have been considered because of their relevant roles in the feedback control mechanisms governing this pathway. In particular, by exploiting the efficiency of tau leaping algorithm, we have investigated the dynamics of the pivotal protein involved in chemotaxis, CheYp, under different conditions, such as the deletion of other chemotactic proteins, the addition of distinct amounts of external ligand, the effect of different methylation states of the receptors.

Concerning the analysis of the interplay between CheYp fluctuations and the number of flagella, other relevant biological aspects of chemotaxis, that stand downstream of the signaling process, might represent valuable points to be considered for future research. For instance, it is known that each flagellar motor switch-complex is constituted by two principal components: a group of proteins, called FliM, FliN, FliG (assembled in a ring), and the torque-generating complexes, called MotA and MotB. In \emph{E. coli}, a typical flagellar ring contains 34 copies of FliM, each of which can bind one copy of CheYp. In \cite{Dyer} it is suggested that binding of CheYp to FliM modifies the displacement of protein FliG, which directly interacts with the Mot complexes and therefore modulates the switch state of the flagellum. Moreover, flagellar motor switching has been found to be highly sensitive to the concentration of CheYp (having a Hill coefficient $\approx$ 10), though the binding of CheYp to FliM has a low level of cooperativity (Hill coefficient $\approx$ 1). In \cite{Dyer}, the hypothesis that CheYp can interact more favorably with the FliM displaced in the CW orientation, than those in the CCW orientation, is put forward. In \cite{Scharf}, in addition, the following mechanism is considered for the control of flagellar motor by means of CheYp: the number of CheYp molecules bound to FliM determines the probability of CW or CCW rotation, while the switch is thrown by thermal fluctuations. In other words, CheYp only changes the stabilities of the two rotational states, by shifting the energy level of CCW-state up and of CW-state down, by a magnitude that is directly proportional to the number of bound molecules. Therefore, interesting questions related to stochastic fluctuations of CheYp, that might be coupled with the investigation on the number of cellular flagella, are: how many FliM proteins at each flagellar switch have to be occupied by CheYp in order to generate the CW rotation of each flagellum and of the bacterium \cite{Bren}? What is the corresponding probability of throwing the reversal switch? Can a double-threshold crossing mechanism \cite{Bren} be more suitable to effectively detect the CW and CCW rotational states?

Nonetheless, other related features might be relevant points for a further extension of this work. For instance, the gradient of CheYp that can be present inside the cytoplasm -- due to the diffusion from the area of its phosphorylation (close to chemotactic receptors) to the area of its activity (around the flagellar motors) -- can be a significant aspect in chemotaxis, together with the localization of CheZ (that controls the dephosphorylation of CheYp) and of the flagella around the cell \cite{Lipkow}. With respect to this matter, how are diffusion processes related to the interactions between CheYp and the flagellar motors? And how does diffusion intervene on the chemotactic response?

We believe that the definition of detailed mechanistic models, like the one proposed in this paper for chemotaxis, coupled to the use of efficient procedures for the analysis of stochastic processes in individual cells, can be a good benchmark to investigate the combined roles of many  biological factors interplaying within a common system. With this perspective, the development of formal methods specifically devised for the analysis of properties (e.g., synchronization) of stochastic systems represents indeed a hot topic research in biological modeling.

\bibliographystyle{eptcs}
\bibliography{biblio}

\end{document}